\documentclass[%
 aip,
 apl,%
 amsmath,amssymb,
 reprint,%
]{revtex4-1}

\usepackage{graphicx}% Include figure files
\usepackage{dcolumn}% Align table columns on decimal point
\usepackage{bm}% bold math
\usepackage[utf8]{inputenc}
\usepackage[T1]{fontenc}
\usepackage{mathptmx}

\begin{document}

\preprint{AIP/123-QED}

%\title{Nitrogen-Vacancy ensembles in a diamond on tapered ultra-high NA optical fiber tip for fluidic environment}
\title{Tapered ultra-high Numerical Aperture optical fiber tip for Nitrogen-Vacancy ensembles based endoscope in a fluidic environment}
% Force line breaks with \\

\author{Dewen Duan}
 \email{dduan@gwdg.de.}%Lines break automatically or can be forced with \\
\author{Vinaya Kumar Kavatamane}%
\author{Sri Ranjini Arumugam}%
\affiliation{ 
Max-Planck Research Group Nanoscale Spin Imaging, Max Planck Institute for Biophysical Chemistry,Am Fassberg 11,G\"{o}ttingen, 37077, Germany%\\This line break forced with \textbackslash\textbackslash
}%
\author{Yan-Kai Tzeng}
\affiliation{Institute of Atomic and Molecular Sciences, Academia Sinica, Taipei 106, Taiwan}
\affiliation{Department of Physics, Stanford University, Stanford, California 94305, USA.}
\author{Huan-Cheng Chang}
\affiliation{Institute of Atomic and Molecular Sciences, Academia Sinica, Taipei 106, Taiwan}
\author{Gopalakrishnan Balasubramanian}
 \email{gbalasu@gwdg.de}
\affiliation{ 
Max-Planck Research Group Nanoscale Spin Imaging, Max Planck Institute for Biophysical Chemistry,Am Fassberg 11,G\"{o}ttingen, 37077, Germany%\\This line break forced with \textbackslash\textbackslash
}%

\date{\today}% It is always \today, today,

\begin{abstract}
Fixing a diamond containing a high density of Nitrogen-Vacancy (NV) center ensembles on the apex of a multimode optical fiber (MMF) extends the applications of NV-based endoscope sensors. Replacing the normal MMF with a tapered MMF (MMF-taper) has enhanced the fluorescence (FL) collection efficiency from the diamond and achieved a high spatial resolution NV-based endoscope. The MMF-taper's high FL collection efficiency is the direct result of multiple internal reflections in the tapered region caused by silica, which has a higher refractive index (RI) than the surrounding air. However, for applications involving fluidic environments whose RI is close to or higher than that of the silica, the MMF-taper loses its FL collection significantly. Here, to overcome this challenge, we replaced the MMF-taper with an ultra-high numerical aperture (NA) microstructured optical fiber (MOF) which is tapered, and sealed its air capillaries at the tapered end. Since the end-sealed air capillaries along the tapered MOF (MOF-taper) have isolated the MOF core from the surrounding medium, the core retains its high FL collection and NV excitation efficiency in liquids regardless of their RI values. Such a versatile NV-based endoscope could potentially find broad applications in fluidic environments where many biological processes and chemical reactions occur. 
\end{abstract}

\maketitle
Nitrogen-Vacancy (NV) center in diamond is a promising quantum sensor for various physical quantities such as magnetic field\cite{Wolf2015,Fedotov2014s,Fedotov2014o}, electric field \cite{Dolde2014}, and temperature\cite{Fedotov2014a,Hayashi2018,Zhang2019}. In traditional NV-based sensing, cumbersome optics such as optical lenses and dichroic mirrors are routinely used. The resulting large size of the setup and the limited focal length of the optical lens have restricted the applications of such NV-based sensors. By grafting a NV-rich diamond on the apex of a multimode optical fiber (MMF), and simultaneously using the same MMF for both exciting NV centers and collecting their fluorescence (FL) is shown to enhance the resulting endoscope-type sensor's compactness, reliability, and flexibility for sensing magnetic field or temperature \cite{Fedotov2014s, Fedotov2014o, Fedotov2014a}. However, the inferior FL collection efficiency of normal MMF necessitates a large NV-rich diamond to be fixed on its apex to achieve optimum FL collection. As a result, the spatial resolution of such an endoscope is often low (larger than 100 \(\mu\)m\cite{Fedotov2014s, Fedotov2014o}). Replacing the MMF with either an ultra-high NA microstructure optical fiber (MOF) or a tapered multimode optical fiber (MMF-taper) tip, or integrating the diamond with a micro-concave mirror reflector has greatly enhanced the spatial resolution of such an endoscope to even less than 10 \(\mu\)m \cite{Fedotov2016o, Duan2018, Dong2018, Duan2019}. While such approaches have certainly improved the spatial resolution, the abilities to employ these endoscope-type sensors in varied sample environments such as liquids are greatly impaired.  This is because, as many liquids have refractive index (RI) values which are close to or higher than that of fiber's constituent material silica, the liquids inflating into the micro-capillaries of the MOF will degrade the qualities of the optical waveguide and diminish its ultra-high NA. In the case of MMF-taper, the liquids surrounding the MMF-taper will reduce the multiple internal reflections. Therefore, both the ultra-high NA MOF and even the MMF-taper cannot achieve optimum FL collection efficiency from the diamond attached to their apexes when immersed inside the liquid samples. 

Here, to overcome this problem, we tapered an ultra-high NA MOF and sealed its air capillaries at its tip to replace the MMF-taper in a NV-based endoscope. Since the end-sealed air capillaries along the MOF-taper isolated the tapered fiber core from the surrounding medium, the high RI difference between the MOF-taper's core and the air capillaries remains unchanged. Thus, the MOF-taper keeps its high FL collection efficiency unaffected by liquids irrespective of their RI. We discuss the relationship between the length of capillaries sealing region, diameter of the MOF-taper's core, and its FL collection efficiency. We also compared NV excitation and FL collection capability of end-sealed MOF and MOF-taper with other fiber tips.

\begin{figure}
\includegraphics[width=8cm]{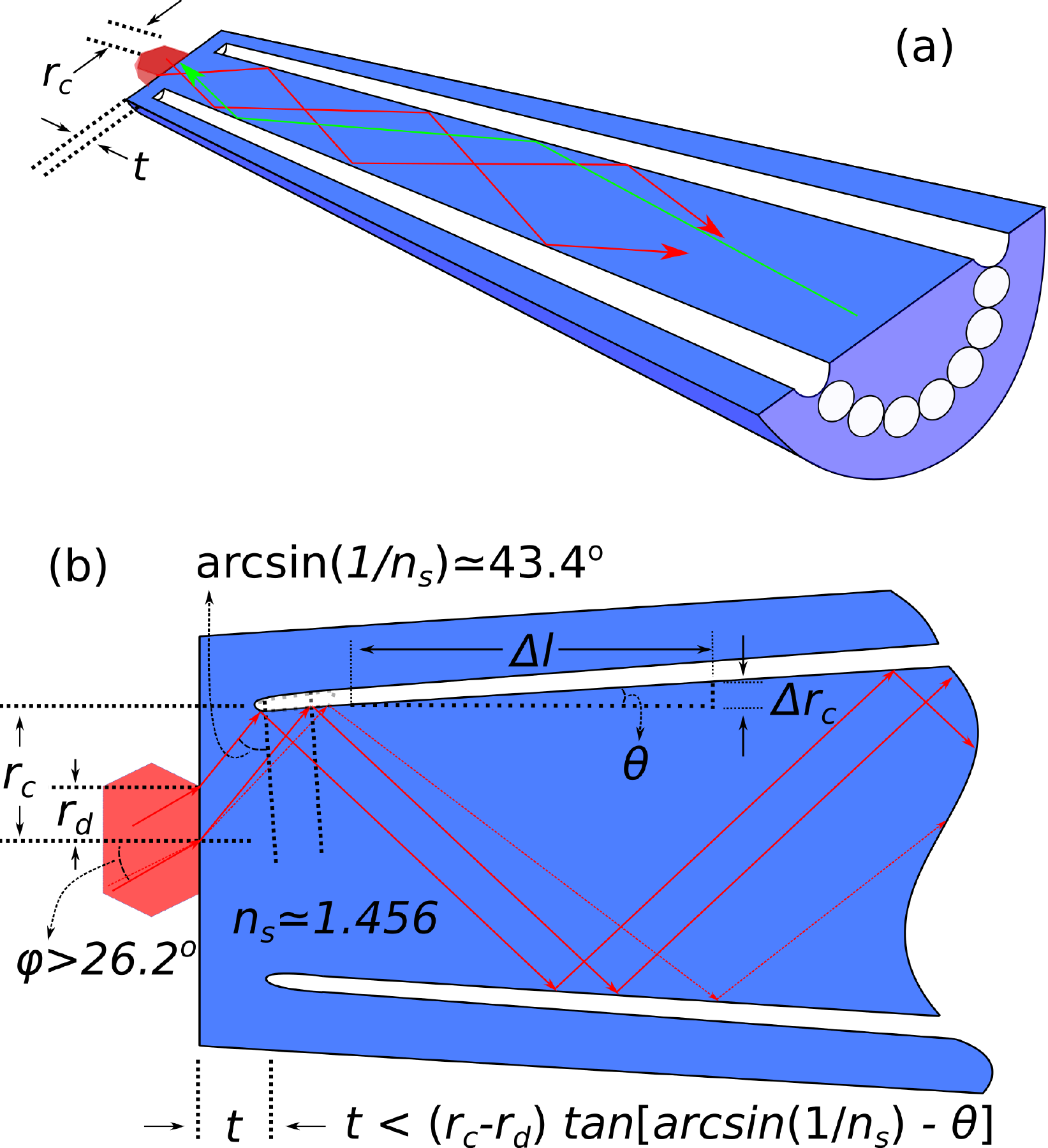}    
\caption{\label{fig: fig.1}The principle schematic of using air capillaries end-sealed ultra-high numerical aperture (NA) microstructure optical fiber (MOF) (tapered or un-tapered) for NV excitation and fluorescence (FL) collection. (a) is a 3-D view of the configuration. (b) is the sectional view and parameter relationship used in Eq.\ref{eq.1}. In (b), if the MOF is not tapered, $ \Delta r_{c} = 0$ and $\theta = arctan(\,\Delta r_{c}/\Delta l) = 0$.} 
\end{figure}

The principle schematic of using end-sealed MOF-taper or MOF for NV excitation and FL collection is shown in Fig.\ref{fig: fig.1}. In an ideal condition (Fig.\ref{fig: fig.1} (b), ignoring the FL from diamond which first enters the air and then gets coupled to the core of MOF-taper or MOF), when the length of the sealed region ($t$) meets the following critical condition, the sealing of air capillaries will not lead to any reduction in FL collection efficiency.
\begin{eqnarray}
   t \leq ( \,r_{c}-r_{d}) \, tan[ \,arcsin(\,1/n_{s})\,-arctan(\,\Delta r_{c}/\Delta l)\, ]\,
\label{eq.1}.
\end{eqnarray}

where $r_{c}$  and $ r_{d}$ are the radii of cores of MOF-taper or MOF and the diamond, respectively. $n_{s} \approx 1.456$ is the RI of silica, $\Delta r_{c}$ is the change in radius (due to tapering) with a corresponding change in length $\Delta l$ and the ratio ($\Delta r_{c}/\Delta l$) defines the taper sharpness. In the case of MOF, a length ($t$) of the sealed region is less than 0.94 times the radius difference of the fiber core and diamond  ( $\Delta r_{c}=0$, $t \leq 0.94(\,r_{c}-r_{d})\,$) guarantees that the critical condition is met and the air capillaries sealing will not reduce its FL collection efficiency. 
\begin{figure}
   \includegraphics[width=8cm]{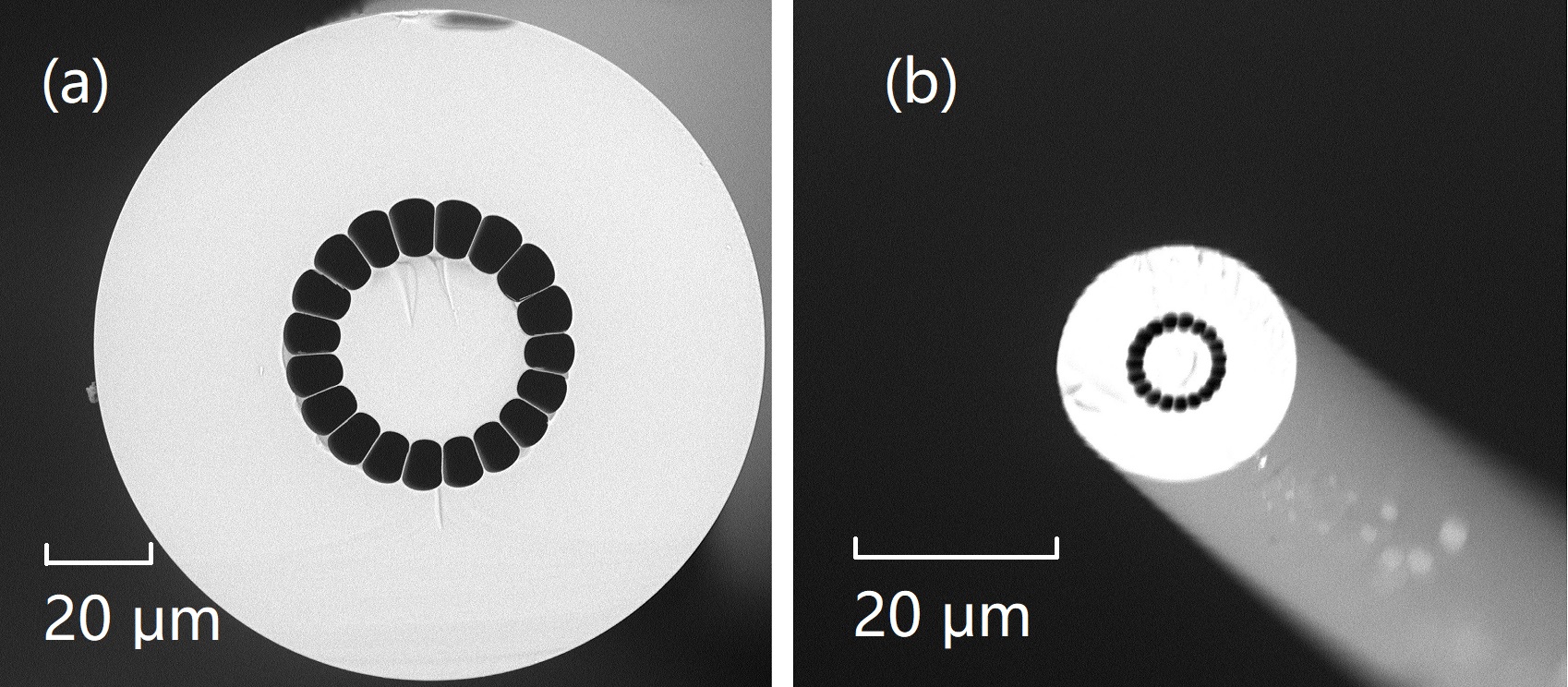}
   
   \includegraphics[width=8cm]{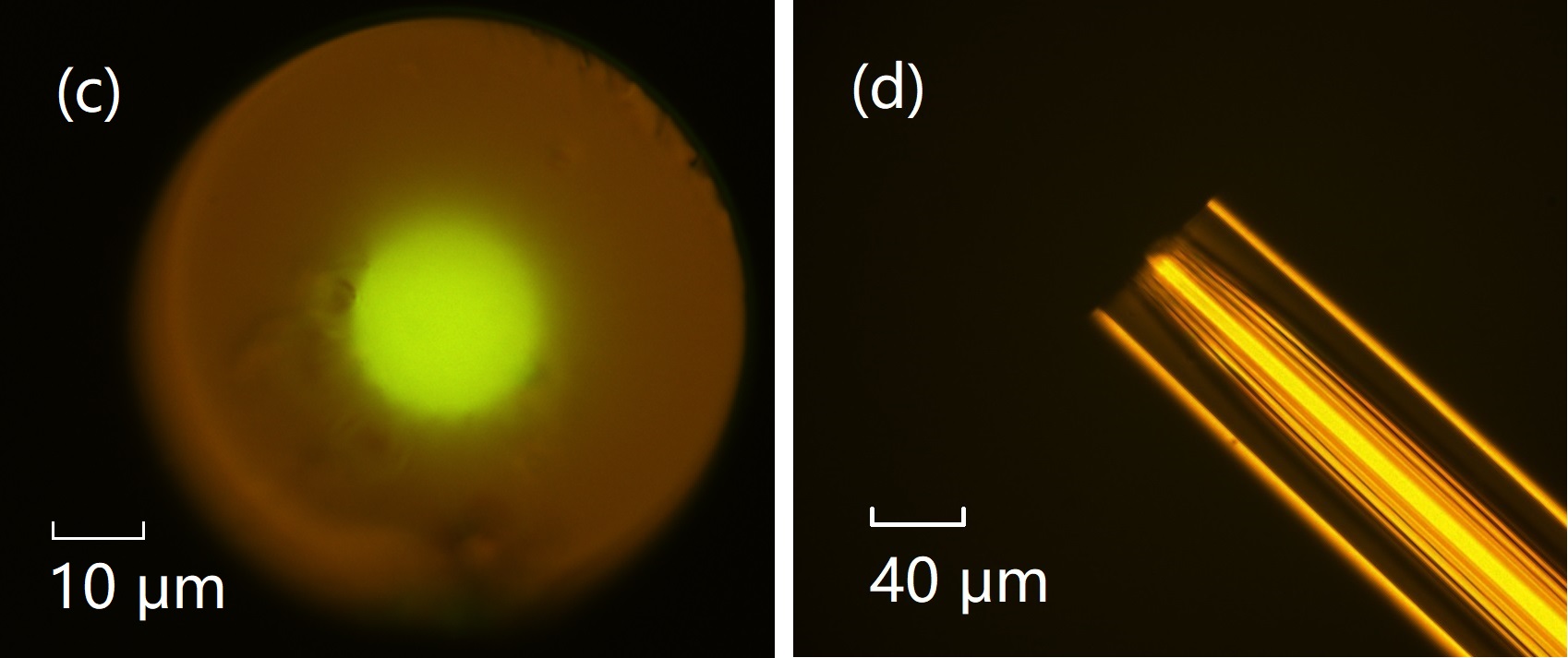}
   % \includegraphics[width=3.9cm]{Fig2_a.jpg}\hspace*{-0.5em}
   % \qquad
   % \includegraphics[width=3.9cm]{Fig2_b.jpg}\hspace*{-0.5em}
    \caption{\label{fig: fig.2} (a) and (b) are scanning electron microscope images of cross-section of the MOF and a MOF taper (tip core diameter is 5.2\(\mu\)m). (c) and (d) are optical microscope images of the cross-section view and side view of a end sealed MOF-taper, respectively.} 
\end{figure}

\begin{figure} 
\includegraphics[width=8cm]{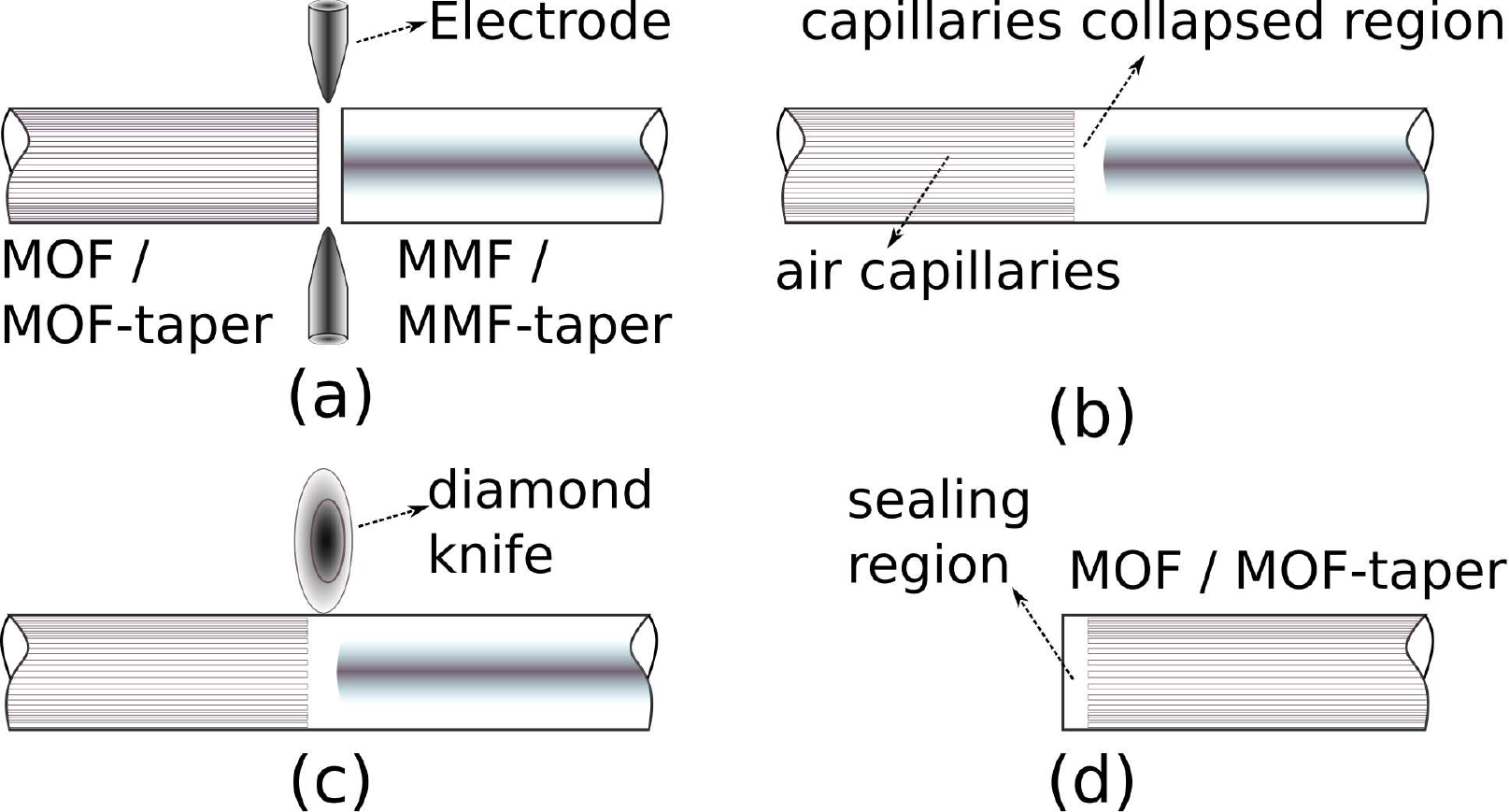}     
\caption{\label{fig: fig.3} Schematic of air capillaries sealing process. (a) Fusion splicing the MMF-taper and MOF-taper (or MMF and MOF) together. (b) is the result of (a). (c) Cleaving the fiber near the spliced joint with a short section of capillaries collapsed region remains on the MOF-taper or MOF; (d) is the result of (c).} 
\end{figure}

The ultra-high NA MOF we used is MM-HNA-35 (NKT photonics; core and cladding diameters are 35$\pm2$ \(\mu\)m and 125$\pm3$ \(\mu\)m respectively) and has a NA larger than 0.6 at 880 nm. Fig.\ref{fig: fig.2} (a) shows its cross-section image. The MOF-taper was fabricated by first stretching the MOF while heating it on flame, and then cutting it at the taper waist. Fig.\ref{fig: fig.2} (b) shows the end-face of a fabricated MOF-taper. The end-sealed MOF-taper (or MOF) tips are fabricated by first fusion splicing the MOF-taper (or MOF) to a section of MMF-taper (or MMF), and then cleaving the MOF-taper (or MOF) near the collapsing region of air capillaries. Optical microscope images of an end-sealed MOF-taper is shown in Fig.\ref{fig: fig.2}(c) and (d). The sealing process is schematically shown in Fig.\ref{fig: fig.3}. An ideal sealing is the one which leaves a very short section of the sealing region (i.e. small \textit{t}) without gradually collapsing of the air capillaries. In the sealing process employed in this work, relatively long air capillaries with a gradually collapsing region would partly ruin the ultra-high NA of the MOF-taper or MOF. 

\begin{figure} 
\includegraphics[width=8cm]{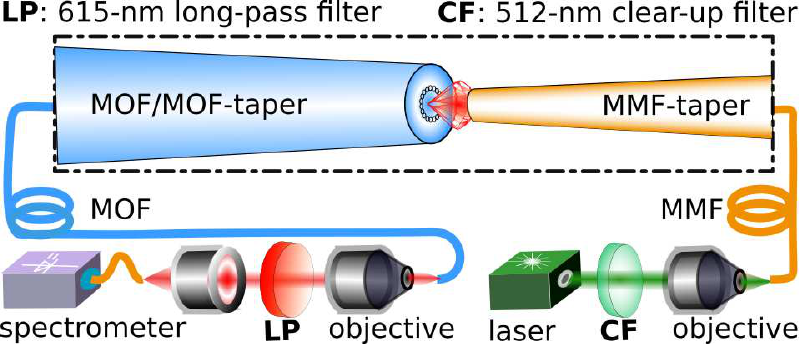}    
\caption{\label{fig: fig.4}Schematic of the experimental setup used to measure NV excitation and FL collection efficiencies of different fiber tips. For measuring the performance of MMF the MOF/MOF taper in the diagram was replaced with MMF/MMF-taper.} 
\end{figure}

\begin{figure} 
\includegraphics[width=8.3cm]{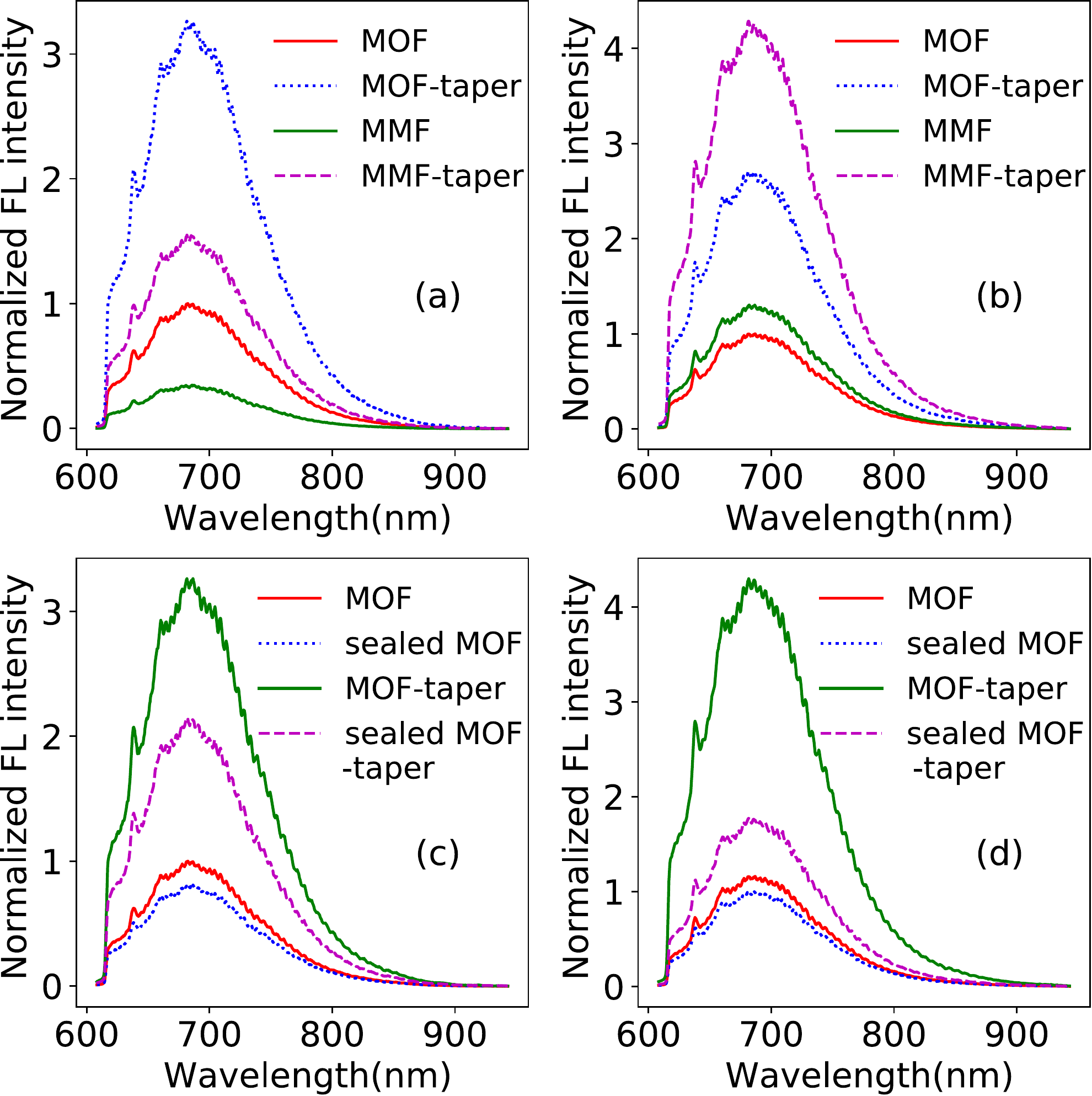}    
\caption{\label{fig: fig.5}FL collection and excitation efficiency measurements. (a) and (c) are the measured FL collection abilities of different fiber tips. They were measured when a $\sim$8 \(\mu \)m diamond on the MMF-taper tip is illuminated with a CW green laser of constant power ($\sim$5.86 mW) in the setup shown in Fig.\ref{fig: fig.4}. (b) and (d) are the measured excitation abilities of different fiber tips. Note that excitation measurements in (b) and (d) are done by letting MMF-taper to collect the FL while a laser with constant power illuminates the diamonds attached to apexes of different fibers.} 
\end{figure}

We first compared the performances of NV excitation and FL collection of MOF-taper (unsealed) with MMF, MOF (unsealed), and MMF-taper, respectively. The experimental setup used to assess their performance is shown in Fig.\ref{fig: fig.4}, and the obtained results are shown in Fig.\ref{fig: fig.5}. The MMF used is GIF625 graded-index fiber, and the MMF-taper tip has a diameter of $\sim$10.2 \(\mu \)m which we have not optimized for collecting the FL from the diamond having a size of $\sim$8 \(\mu \)m. The MOF-taper used in the test is with a tip core diameter of $\sim$13.2 \(\mu \)m.  The results in Fig.\ref{fig: fig.5} (a) show that the MMF-taper has $\sim$4.5 times FL collection efficiency of the un-tapered MMF (can be further increased by optimizing the MMF-taper \cite{Duan2019}). On the other hand, tapering a MOF achieves over $\sim$3 times the FL collection efficiency of the original un-tapered MOF. It is also evident that overall FL collection of MOF is  $\sim$3 times that of MMF. These observations are reasonable since the tapering of both MOF and MMF significantly increases the NA of tapered tips, and MOF has a much higher NA than that of MMF. Fig.\ref{fig: fig.5}(b) shows the NV excitation efficiency of MOF and MMF. While a tapered MMF shows $\sim$ 3.35 times improvement in NV excitation efficiency compared to bare MMF, a tapered MOF results in relatively lower ($\sim$2.75 times) improvement in NV excitation efficiency. We believe that this difference in excitation efficiencies between MMF and MOF is due to the fact that MOF-taper has a larger core size than the MMF-taper tip. Fig.\ref{fig: fig.5}(c) and (d) show that sealing the ends of both the MOF and MOF-taper using the procedure described in Fig.\ref{fig: fig.3} significantly degrades their FL collection and NV excitation efficiency. Sealing the MOF-taper found to reduce its FL collection as much as $\sim$ 34.2\% and NV excitation efficiency by  $\sim$ 59.4\% . While sealing the un-tapered MOF had relatively lower influence with FL collection efficiency dropping by $\sim$ 19.2\%
and NV excitation efficiency by $\sim$ 14.2\%. According to Eq.\ref{eq.1}, when collecting the FL from a diamond of $\sim$8 \(\mu \)m diameter, if the sealed region of the MOF is shorter than $\sim$12.7 \(\mu \)m ($\sim$0.94(35-8)/2), the reduction in FL collection efficiency induced by sealing the tip can be ignored. As a comparison, the sealed region should be less than $\sim$2.4 \(\mu \)m ($\sim$0.94(13.2-8)/2) for the MOF-taper to guarantee that the tip sealing does not cause any reduction in FL collection efficiency. Obviously the sealing process described in Fig.\ref{fig: fig.3} is unable to achieve such a short sealed region. Hence, a more advanced sealing technique which produces better results is recommended.

\begin{figure} 
\includegraphics[width=8.3cm]{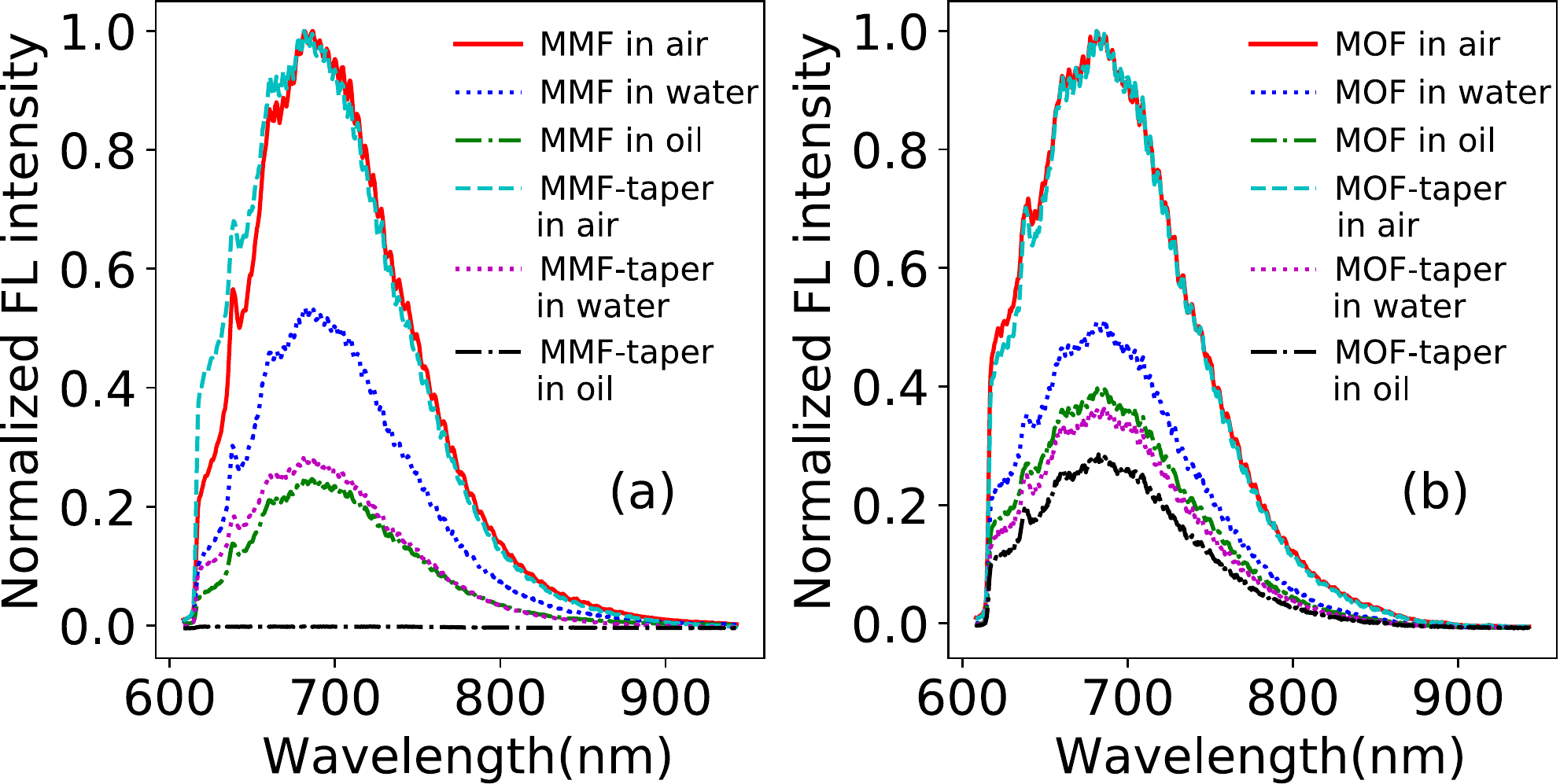}    
\caption{\label{fig: fig.6}Performances of the different fiber tips in the liquid environments. (a) Diamond on MMF and MMF-taper tips in air, water, and oil media. The sizes of diamond crystals are $\sim$48-\(\mu \)m and $\sim$8.0-\(\mu \)m on MMF and MMF-taper tips, respectively. (b) Diamonds on air capillaries end-sealed MOF and MOF-taper tips in air, water, and oil. The diamonds' sizes are $\sim$13.7 \(\mu \)m and $\sim$8.1 \(\mu \)m for MOF and MOF-taper, respectively. All collected FL intensity is normalized by the peak values in air and the incident laser power was kept constant for each case.} 
\end{figure}

For studying the effects of liquid environment on the FL collection efficiencies we fixed diamond crystals on the tips of MMF, MMF-taper, sealed MOF and sealed MOF-taper. Fig.\ref{fig: fig.6} summarizes
the performances of different fibers containing diamonds
in water (RI $ n_{w}= 1.33 $) and oil (RI $ n_{o}= 1.51 $; microscopy immersion oil 51786, Fluka.) media. The data are normalized with the
corresponding values from the diamond in air. Fig.\ref{fig: fig.6} (a) shows
that for the case of MMF containing $\sim$48-\(\mu \)m diamond on its
apex, FL losses of $\sim$46.5\% and $\sim$75.4\% was observed in water
and oil, respectively. For MMF-taper with a $\sim$8.0-\(\mu \)m diamond,
$\sim$71.8\% of FL loss when in water, and an almost 100\% FL loss in
the oil was observed (black dotted line in Fig.\ref{fig: fig.6} (a)). On the other hand Fig.\ref{fig: fig.6} (b) indicates that for an end-sealed MOF-taper attached to a $\sim$8.1-\(\mu \)m diamond, the FL losses were
found to be $\sim$63.8\% in water and $\sim$71.5\% in oil.

\begin{figure} 
\includegraphics[width=8.3cm]{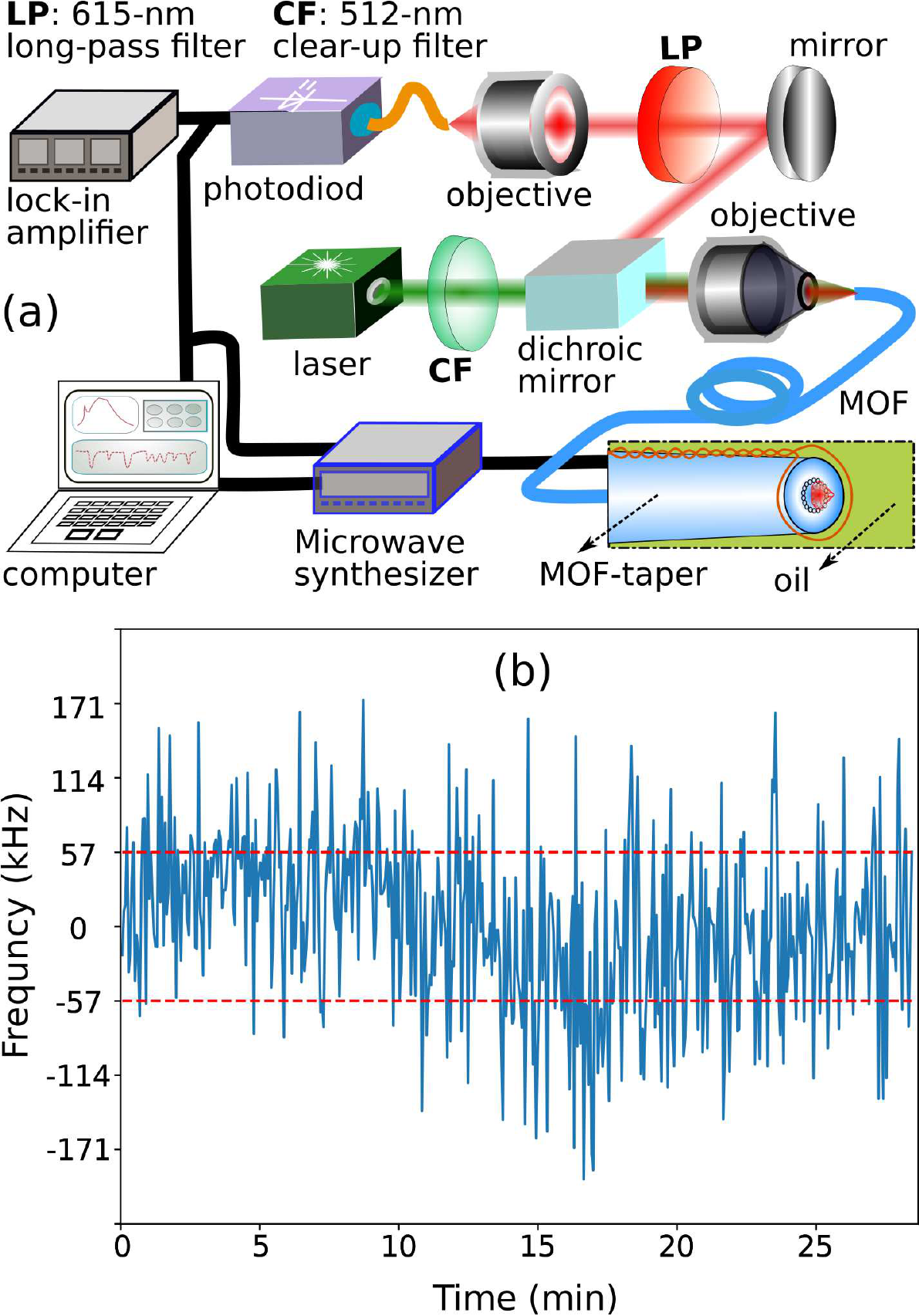}    
\caption{\label{fig: fig.7} (a) Experimental setup for evaluating the performance of MOF-taper and the NV based endoscope. (b) The tested ODMR signal stability of NV based endoscope with MOF-taper.} 
\end{figure}

It is clear that the loss of collected FL in  Fig.\ref{fig: fig.6} is different for different fiber tips. When inserted into liquids, all these fiber tips (MMF, MMF-taper, MOF, and MOF-taper) suffer from the reduction of FL entering into these tips. In addition, due to the reduced RI difference at the diamond-liquid interface and hence the reduced multiple internal reflections cause inefficient NV excitation by the green laser as well as poor coupling of emitted FL into the fiber. For the MMF-taper, there is an additional FL loss caused due to reduction of multiple internal reflections when the liquid’s RI approaches that of MMF-taper. As evident from the case of MMF-taper in oil (Fig.\ref{fig: fig.6}(a)), when RI of surrounding medium was equal or slightly higher (n = 1.51) than RI of MMF-taper, no multiple internal reflections exist inside the MMF-taper and as a result, the MMF-taper collects almost no FL. Therefore, it is obvious that even though the MMF-taper has a high NA in air, it is not suitable for working with liquids whose RI values are close to or higher than the RI of silica. However, from Fig.\ref{fig: fig.6}(b) we can see that the MOF-taper offers much better FL collection than MMF-taper when inserted in oil. Importantly, the sealed MOF-taper offers a high spatial resolution (due to small diamonds) while keeping a relatively high fluorescence collection efficiency in liquids with RI values close to or higher than that of optical fiber material. It is true that even in an ideal condition (which satisfies the Eq.\ref{eq.1}), both sealed MOF and sealed MOF-taper suffer from losses in FL collection due to reduced internal reflections inside the diamond, just like in the case of MMF. However, when the sealed region is not short enough, both MOF and MOF-taper will suffer an extra loss in FL collection which escapes into surrounding liquid. As the sealed MOF and MOF-taper fabricated by us have sealed region length (\textit{t}) longer than the critical value, it is not surprising that there is an extra FL loss when the sealed MOF and MOF-taper are inserted into water and oil (Fig.\ref{fig: fig.6}(b)). Using advanced sealing techniques to shorten the sealing region will further enhance the FL collection efficiency of the sealed MOF-taper.

Finally, we fixed a $\sim$8.6-\(\mu \)m diamond on an air capillaries end-sealed MOF-taper apex using UV-curing glue and tested its magnetic field sensing performance in the oil. We used the experimental setup described in Fig.\ref{fig: fig.7} (a) and a result from optically detected magentic resonance (ODMR) is shown in Fig.\ref{fig: fig.7} (b). The result shows that the ODMR demodulation can be stabilized to 0.3 MHz (in a 2.7 mT field environment), that is $\sim$10.7 \(\mu \)T (obtained by dividing 0.3 MHz by 28 MHz/mT).

In conclusion, we have proposed an ultra-high NA MOF-taper for the NV-based endoscope. The MOF-taper is fabricated by first tapering the MOF and then sealing its micro air capillaries at the tapered tip. Our MOF-taper holds a relatively high fluorescence collection efficiency in liquids with any refractive index values. We discussed the dependence of fluorescence collection efficiency on the sealing length of micro-capillaries, and core size of the MOF-taper. We also experimentally compared the performance of fabricated MOF-taper with other optical fiber tips, and the results demonstrated that MOF-taper holds a relatively high  fluorescence collection efficiency in immersion oil with a RI value as high as n = 1.51. Such a NV-based endoscope could potentially find applications in fluids samples relevant in several chemical and biological studies.

This research is sopported by the  Max-Planck-Gesellschaft (MPG) and Niedersächsische Ministerium für Wissenschaft und Kultur.

% If in two-column mode, this environment will change to single-column format so that long equations can be displayed. 
%\begin{widetext}
%$$\mbox{put long equation here}$$
%\end{widetext}

% Figures should be put into the text as floats. 
% Use the graphics or graphicx packages (distributed with LaTeX2e).
% Put the label that you will use with \ref{} command in the braces of the \label{} command.
%
% \begin{figure}
% \end{figure}

% \begin{table}
% \end{tabular}
% \end{table}

%\begin{acknowledgments}
% Put your acknowledgments here.
%\end{acknowledgments}

\bibliography{aiptemplate}

\end{document}